\documentclass[twocolumn,aps,prd,floatfix,preprintnumbers,a4paper,nofootinbib,superscriptaddress,10pt]{revtex4-1}

\usepackage{epsfig,graphics,graphicx}
\usepackage{amsmath,amssymb,mathtools,amsfonts}
\usepackage{bm,soul}
\usepackage[usenames,dvipsnames]{color}
\usepackage{amssymb,url,psfrag,times}
\usepackage[varg]{txfonts}
\usepackage[colorlinks, pdfborder={0 0 0}]{hyperref}
\usepackage{lineno,verbatim}
\usepackage[utf8]{inputenc}
\usepackage{ulem}
\usepackage{xcolor}

\usepackage[english]{babel}
\usepackage{blindtext} 

\definecolor{LinkColor}{rgb}{0.75, 0, 0}
\definecolor{CiteColor}{rgb}{0.75, 0, 0}
\definecolor{UrlColor}{rgb}{0.75, 0, 0}

\hypersetup{linkcolor=LinkColor}
\hypersetup{citecolor=CiteColor}
\hypersetup{urlcolor=UrlColor}
\def\check#1{#1}

\usepackage{inputenc}
\usepackage{pgfplots}
\usepackage{tikz}
\usetikzlibrary{shapes,arrows}
\usetikzlibrary{positioning}
\usetikzlibrary{backgrounds}

\newcommand*{\figfactor}{0.495}

\newcommand{\mlam}{{\bm{\lambda}}}

\definecolor{lightblue}{rgb}{.82,.88,0.95}
\definecolor{lightred}{rgb}{0.95,.86,0.86}
\definecolor{yellow}{rgb}{0.95,0.95,0.86}
\definecolor{green}{rgb}{.90,1,0.95}
\definecolor{lightpurple}{rgb}{.95,0.85,0.95}

\tikzset{
    vertex/.style = {
        circle,
        fill            = black,
        outer sep = 2pt,
        inner sep = 1pt,
    }
}

\def\gt{Georgia Tech}
\def\tk{Teukolsky}

\def\grad#1{gravitational radiation#1}
\def\gr#1{general relativity#1
  (GR#1)\gdef\gr{GR}}

\def\gw#1{gravitational wave#1}

\def\nht#1{no-hair hypothesis#1}

\def\tk#1{Teukolsky#1}

\def\fig#1{Figure~\ref{#1}}

\def\eqn#1{Equation~(\ref{#1})}

\newcommand{\eqns}[2]{Eqs.~(\ref{#1})-(\ref{#2})}

\def\lal#1{LIGO Analysis Library#1
  (LAL#1)\gdef\lal{LAL}}
\def\nrda#1{\nr{} Data Analysis#1
  (NRDA#1)\gdef\nrda{NRDA}}
\def\tt#1{\textit{transverse--traceless}#1
  (TT#1)\gdef\tt{TT}}
\def\et#1{Einstein Telescope#1
  (ET#1)\gdef\et{ET}}
\def\ego#1{European Gravitational Observatory#1
  (EGO#1)\gdef\ego{EGO}}
\def\elisa#1{Evolved Laser Interferometer Space Antenna#1
  (eLISA#1)\gdef\elisa{eLISA}}
\def\ligo#1{Laser Interferometer Gravitational Wave Observatory#1
  (LIGO#1)\gdef\ligo{LIGO}}
\def\virgo#1{Virgo#1}
\def\aligo#1{advanced LIGO#1
  (Adv. LIGO#1)\gdef\aligo{Adv. LIGO}}
\def\snr#1{signal-to-noise ratio#1
  (SNR#1)\gdef\snr{SNR}}
\def\psd#1{power spectral density#1
  (PSD#1)\gdef\psd{PSD}}
\def\rom#1{reduced order model#1
  (ROM#1)\gdef\rom{ROM}}
\def\gatech#1{Georgia Institute of Technology#1
  (GaTech#1)\gdef\gatech{GaTech}}
\def\ffi#1{Fixed-Frequrncy Integration#1
  (FFI#1)\gdef\ffi{FFI}}
\def\sxs#1{Simulating Extreme Spacetimes#1
  (SXS#1)\gdef\sxs{SXS}}
\def\bam#1{Bifunctional Adaptive Mesh#1
  (BAM#1)\gdef\bam{BAM}}
\def\adm#1{Arnowitt-Deser-Misner
	(ADM#1)\gdef\adm{ADM}}
\def\frmse#1{Fractional Root-Mean Square Error
	(FRMSE)\gdef\frmse{FRMSE}}

\def\bh#1{black hole#1
 (BH#1)\gdef\bh{BH}}
\def\bbh#1{binary black hole#1
 (BBH#1)\gdef\bbh{BBH}}

\def\qnm#1{quasinormal mode#1
  (QNM#1)\gdef\qnm{QNM}}
\def\eob#1{Effective One Body#1
  (EOB#1)\gdef\eob{EOB}}
\def\gw#1{gravitational wave#1}

\def\pn#1{post-Newtonian#1
 (PN#1)\gdef\pn{PN}}
\def\pnl#1{post-Newtonian-like#1
  (PN-like#1)\gdef\pnl{PN-like}}

\def\nr{numerical relativity
 (NR)\gdef\nr{NR}}
\def\pt{\bh{} perturbation theory}

\def\rd{ringdown}
\def\imr{inspiral-merger-ringdown}

\def\bbc#1{binary black hole coalescence#1}

\def\pca#1{principle component analysis#1
  (PCA#1)\gdef\pca{PCA}}
\def\svd#1{Singular Value Decomposition#1
  (SVD#1)\gdef\svd{SVD}}

\newcommand\hidetosubmit[1]{}
\newcommand\optional[1]{}
\newcommand\ForInternalReference[1]{}

\newcommand{\cw}{\tilde{\omega}}

\newcommand{\LL}{\bar{ \ell }}

\newcommand{\Yj}{{_{-2}}Y_{\LL\MM}(\iota,\phi)}

\newcommand{\MM}{\bar{m}}

\newcommand{\Ak}{A_{\LL\MM \ell n}}
\newcommand{\LM}{\LL\MM}
\def\lmn{ \ell \MM n }
\def\hatt{\hat{t}}
\newcommand{\tref}{ \hatt_{\mathrm{ref}} }
\newcommand{\lmtitle}[2]{\vspace{-0.80cm} \begin{center} \noindent\rule{0.35\paperwidth}{0.3pt} \end{center} \vspace{-0.3cm}}

\def\figfactor{0.3333}

\def\T0{20}
\def\NumNRSims{\check{101}}																		
\def\rdnp{{RDNP}}

\begin{document}


\title{Modeling ringdown II. Aligned-spin binary black holes, implications for data analysis and fundamental theory }

\author{L. London}
\affiliation{LIGO Laboratory and MIT-Kavli Institute for Astrophysics and Space Research, 77 Massachusetts Avenue, 37-664H, Cambridge, Massachusetts 02139, USA}
\affiliation{School of Physics and Astronomy, Cardiff University, The Parade, Cardiff, CF24 3AA, United Kingdom}

\begin{abstract}
	The aftermath of binary black hole coalescence is a perturbed remnant whose gravitational radiation rings down, encoding information about the new black hole's recent history and current state.
	It is expected that this ringdown radiation will be composed primarily of Kerr quasinormal modes, and thereby enable tests of general relativity.
	Here, the first complete ringdown signal model for nonprecessing binary black hole systems is presented: multipole amplitudes and phases are modeled as functions of initial binary parameters.
	It is found that using the peak time of the dominant merger multipole as a reference results in the dominant mode's excitation being a remarkably simple linear function of system parameters, strongly suggesting that an analytic treatment may be within reach.
	In particular, for initially nonspinning black holes, the dominant quadrupole is excited as $-4$ times the system's symmetric mass ratio.
	%
	%
	Application of the model to parameter estimation allows general relativity predictions for mode amplitudes independently of signal strength.
	Treatment of GW150914 indicates some mode amplitudes and relative phases are intrinsically difficult to constrain.
	%
	%
\end{abstract}


\date{\today}

\maketitle

\section{Introduction}
%
%
\par Direct detections of gravitational waves by LIGO and \virgo{} bring the possibility of testing \gr{'s} detailed predictions
\cite{Abbott:2017oio,Abbott:2017vtc,TheLIGOScientific:2016src,Yunes:2016jcc,TheLIGOScientific:2016pea}.
With prospective detectors such as LIGO-India~\cite{Unnikrishnan:2013qwa}, KAGRA~\cite{Kanda:2017cwi}, \et{}~\cite{Maselli:2017kvl} and LISA~\cite{Tang:2018rfm}, it is likely that there will be many high signal-to-noise ratio (SNR) detections,
allowing for increasingly stringent tests of \gr{}
\cite{Bhagwat:2016ntk, Baibhav:2017jhs, Berti:2016lat, Yang:2017zxs, Cabero:2017avf}.
To this end, the final moments of \bbc{} are of particular interest.
Shortly after two \bh{s} merge, the remnant is expected to be a perturbed \bh{} whose \grad{} rings down with frequencies predicted by \tk{'s} equations \cite{leaver85,PhysRevLett.29.1114}.
In particular, classical linear perturbations of the Kerr spacetime induce transient radiative \qnm{s} that are exponentially damped and oscillatory\cite{gr-qc/9810074,LivRevQNM,Berti:2009kk}.
The damped ringing of these modes is colloquially named \textit{\rd{}} \cite{gr-qc/9810074}.
\par It is expected that the spatiotemporal dependence of each \qnm{} is determined by the remnant's mass and spin, which in turn determine the matter-free background metric (e.g. \cite{Gurlebeck:2015xpa}).
Consequently, direct observation of two or more \qnm{s} has been linked to testing the \nht{} \cite{Gossan:2011ha,Meidam:2014jpa,Berti:2005ys}.
\par However, significant challenges must first be overcome.
Accurate and physically parametrized signal models are needed to interface theory with experiment.
%
%
Despite the development of \pn{} theory to map initial binary parameters to \gw{}s for the early inspiral, there is no equivalent analytic theory developed for \bh{} \rd{} \cite{Berti:2009kk}.
As a result, \nr{} simulations have been used to provide \qnm{} amplitudes and their relative phases where \pt{} only provides the \qnm{}'s spatiotemporal functions
\cite{Kamaretsos:2011um, Buonanno:2006ui, PhysRevD76Berti, London:2014cma, McWilliams:2018ztb}.
Concurrently, signal models for \bbh{} inspiral, merger and \rd{}, are often limited by their subdominant harmonic content \cite{London:2017bcn,Cotesta:2018fcv}, or are not readily parametrizable for deviations from \pt{}'s predictions \cite{Varma:2018mmi}.
%
%
%
%
\begin{figure}
	\includegraphics[width=0.48\textwidth]{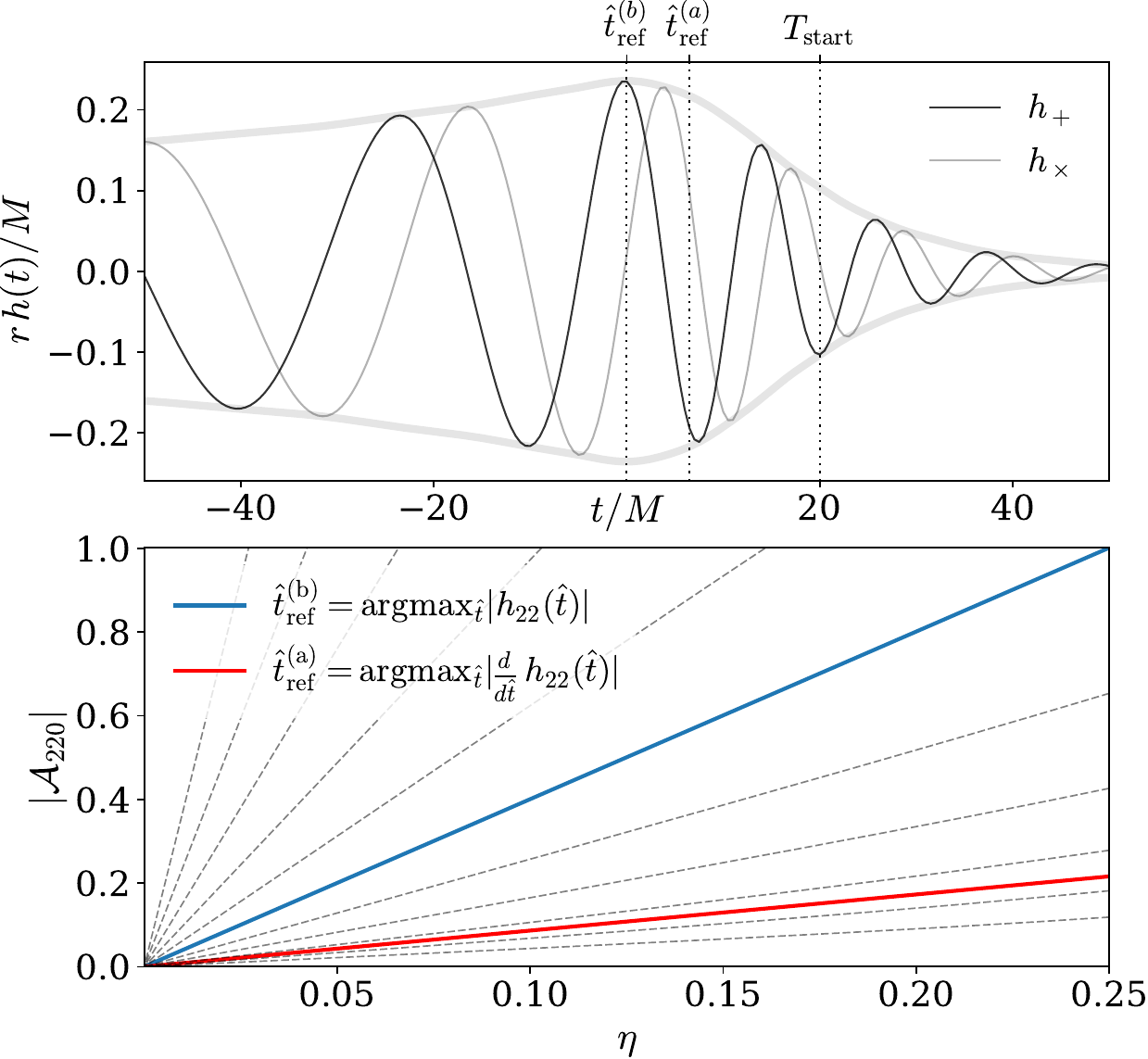}
	\caption{The choice of time origin, $\tref$, affects \rd{} amplitude values independently of \rd{}'s beginning at $t=T_\mathrm{start}$.
	Top panel: Result of \nr{} simulation of 1.2:1 mass ratio \bbh{} with nonspinning progenitors.
	The \gw{} strain is shown at an orientation of $(\theta,\phi)=(14\pi/5,0)$.
	Thick grey curves show its envelope, and vertical dotted lines mark two choices for reference time.
	Bottom panel: the effect of reference time on the dominant \qnm{} amplitude, $\mathcal{A}_{220}$, as a function of symmetric mass ratio, $\eta=M_1 M_2/M^2$.
	Dashed grey curves show the effect of varying $\tref$ between $-25~M$ and $25~M$ in steps of $5~M$ relative to peak strain. Curves shown are approximately linear.
	The use of $\tref^{(a)}$ yields the red curve as used in e.g. Ref.~\cite{Gossan:2011ha}.
	Use of $\tref^{(b)}$ (blue) has the particular effect of normalizing the \qnm{} amplitude to unity when $\eta=0.25$ (i.e. an equal mass binary ).}
	\label{fig:tdfd}
\end{figure}
%
%
%
%
%
%
%
%
\par In this work, the first detailed signal model for \qnm{} excitations (amplitudes and phases) of spinning but nonprecessing \bh{} binaries is presented.
This work is the sequel to, Ref.~\cite{London:2014cma}, which only explores \bbh{}s with nonspinning progenitors.
Because this signal model presented here outputs the expected \rd{} radiation of nonprecessing \bbh{} systems, we will refer to it as \rdnp{}.
\rdnp{}'s construction and output shed new light on the potential development of a \pn{}-like theory for \bh{} \rd{}.
For the first time, \rdnp{} shows that non-monotonic excitations and abrupt transitions in relative phase are robust features of the nonprecessing \bbh{} parameter space.
Its primary use is expected to be in testing \gr{} during and after LIGO's third observing run (O3) \cite{Meidam:2014jpa,Gossan:2011ha,Cabero:2017avf,Yang:2017zxs,Carullo:2018sfu,Carullo:2019flw}.
%
%
%
While there is a focus here on ground based detectors, the primary results of this work apply to proposed space based detectors such as LISA \cite{Berti:2005ys}.
\begin{figure*}[ht]
	\begin{tabular}{ccc}



		%
		\includegraphics[width=\figfactor\textwidth]{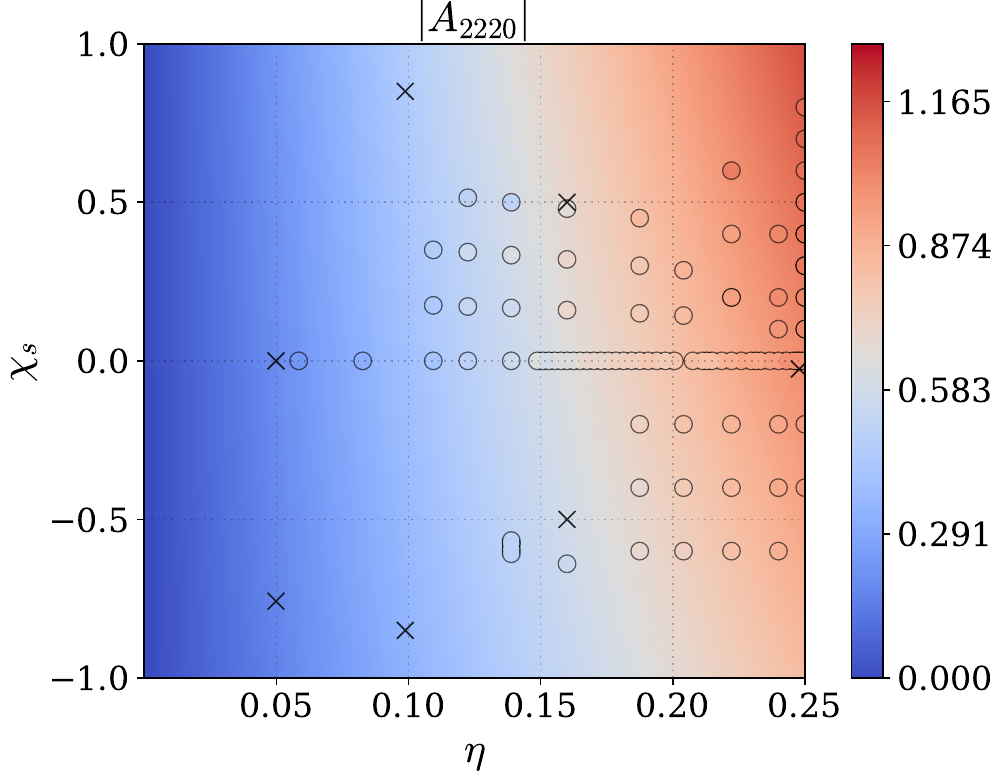} &
		\includegraphics[width=\figfactor\textwidth]{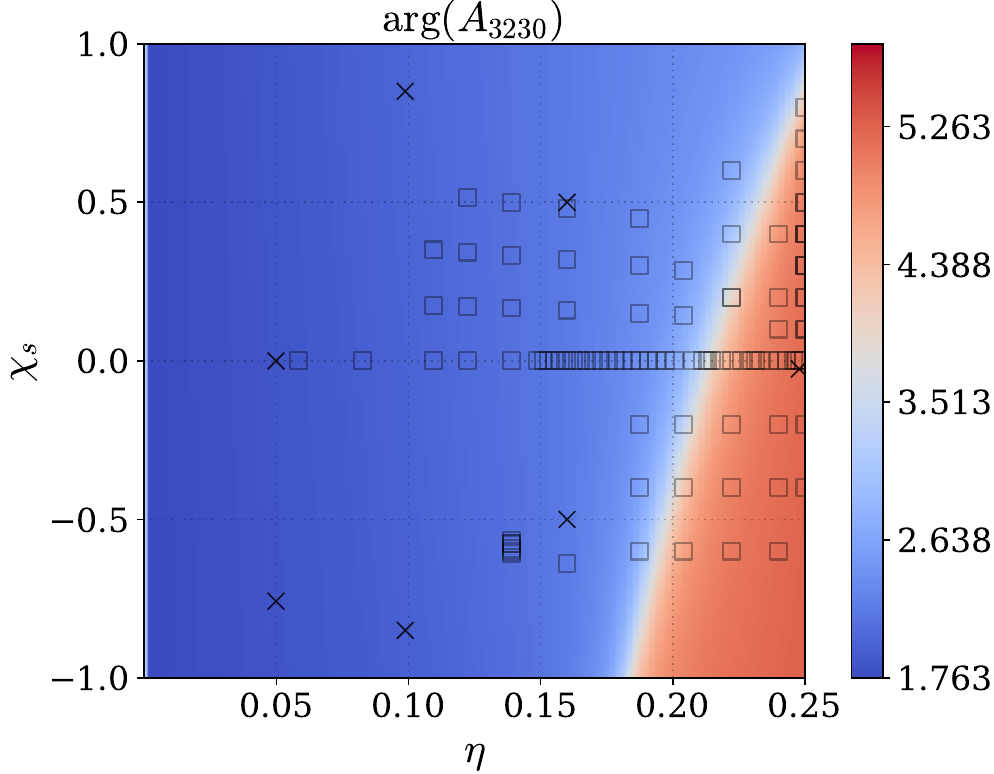} &
		\hspace{-0.0cm}\includegraphics[width=0.318\textwidth]{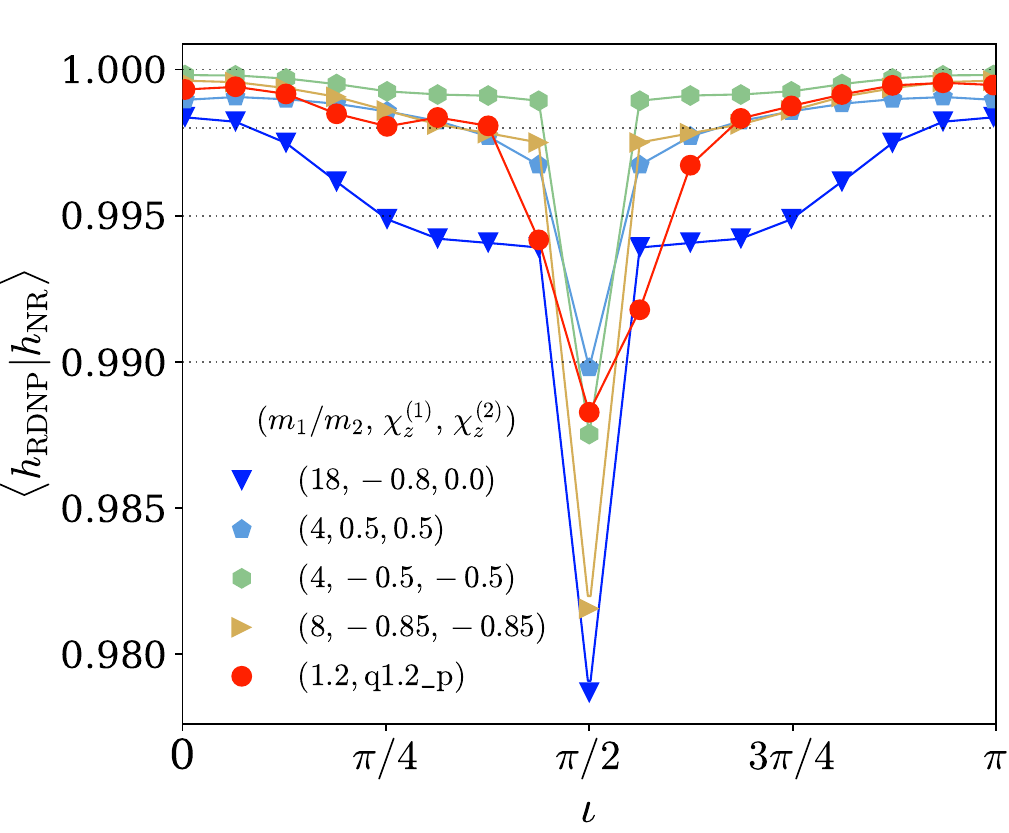}
		%
		%

	\end{tabular}
	\caption{Construction and validation of \rdnp{}:
	(\textit{Left}) 2D surface plot comparing calibration points (colored circles) to model fit (smooth gradient) for $|A_{2220}|$. Values of calibration points differ from model fit if adjacent colors differ. {BAM} validation waveforms are marked with {x}.
	%
	%
	%
	(\textit{Center}) 2D surface plot for the intrinsic phase  $\arg(A_{3230})$. Here, $\delta$ is explicitly considered to be a function of $\eta$.
	(\textit{Right}) Average matches as a function of source inclination for select noncalibration simulations including all multipoles with $\ell \le 5$. Each case has a total system mass of $100~M_{\odot}$.
	A sample precessing system (red circles, label ``q1.2\_p'') having $(\chi^{(1)}_x,\chi^{(1)}_y,\chi^{(1)}_z) = (0.3844,-0.1346,-0.1189)$ and $(\chi^{(2)}_x,\chi^{(2)}_y,\chi^{(2)}_z) = (-0.3536,0.2181,0.0861)$ is shown in addition to 4 nonprecessing cases.}
	%
	%
	\label{fig:results}
\end{figure*}
%
\section{Numerical Relativity Simulations}
%
\par Using \NumNRSims{} nonprecessing simulations from the \gt{} catalog, strain \qnm{} amplitudes are calculated and then modeled in geometric units ($M=G=c=1$)~\cite{Jani:2016wkt}.
Among the simulations used, \check{42} are nonspinning, \check{31} have different dimensionless spins on each \bh{}, and \check{28} have equal spin on each \bh{}.
Mass ratios vary between \check{1:1} and \check{1:15}, and component spins vary between \check{-0.8} and \check{0.8}.
Seven simulations (six nonprecessing and one precessing) from the {BAM} code are used for model validation~\cite{bruegmann:2008,Husa:2007hp}.
\section{Ringdown Start}
%
%
This section reviews the data processing choices used to define \rd{} within simulations of merging \bbh{}s. 
Briefly, the connection between observable \gw{} strain, and the output of \nr{} simulations is discussed.
More importantly, the impact of the extrinsically chosen ringdown start time on ringdown amplitudes is reviewed. 
A choice that simplifies the behavior of the dominant quadrupole amplitude is presented.
\par Given the \gw{} strain, $h = h_+ - i h_\times$, where $h_+$ and $h_\times$ are the observable \gw{} polarizations, a multipolar representation convenient for \nr{} uses the spherical harmonics of spin weight $-2$ ~\cite{Alcubierre.0707.4654}
\begin{align}
	\label{eq:hlm}
	r \, h_{\LL \MM}(t) &= {\int_{\Omega}} {_{-2}}Y_{\LL \MM}^{*}(\theta,\phi)h(t,\theta,\phi) \, \mathrm{d}\Omega
	\; .
\end{align}
Concurrently, \pt{} confers that strain is naturally represented as a sum over the physical system's eigenmodes
\begin{align}
	\label{eq:hlmn}
	r \, h(t,\theta,\phi) \approx \sum_{\ell m n} \, {_{-2}}S_{\ell m n}(\theta,\phi) \,  \mathcal{A}_{\ell m n} e^{ i  \cw_{\lmn} t}
	\; ,
\end{align}
where $\ell \ge 2$, $|m| \le \ell$, and $n \ge 0$.
%
%
In \eqn{eq:hlm}, $r$ is the source's luminosity distance and $*$ denotes complex conjugation.
In \eqn{eq:hlmn}, ${_{-2}}S_{\ell m n}(\theta,\phi)$ is a spheroidal harmonic function, $\cw_{\lmn} = \omega_{\lmn} + i/\tau_{\lmn}$ is the \qnm{}'s complex ringdown frequency, $n$ is an overtone index, and $\mathcal{A}_{\ell m n}$ is the \qnm{} excitation amplitude \cite{leaver85,Berti:2009kk}.
In principle, \eqn{eq:hlmn} is approximate as other possible contributions to the radiation, such as power-law tails, are not included. However, this work focuses on the regime in which \qnm{} decay dominates~\cite{PhysRevLett.29.1114,London:2014cma}. Thus, in practice, the \qnm{} representation is considered to be exact when the start of \rd{} is appropriately chosen, and the self-consistency of this picture well established \cite{London:2014cma,Kelly:2012nd}.
\par The combination of \eqns{eq:hlm}{eq:hlmn} yields that \eqn{eq:hlm}'s spherical harmonic multipole moments are sums over \eqn{eq:hlmn}'s eigenmodes~\cite{Press:1973zz,London:2014cma}
\begin{align}
	\label{eq:hlm2}
	r\,h_{\LL \MM} = \sum_{\ell n} A_{\LM \ell n} e^{ i  \cw_{\lmn} t  }
	\; .
\end{align}
In \eqn{eq:hlm2}, we have used \eqns{eq:hlm}{eq:hlmn} along with the orthogonality of spherical and spheroidal harmonics in $m$.
Moreover, an effective \qnm{} amplitude $	A_{\LM \ell n}$, and a time coordinate, $t$, are defined as
\begin{align}
	\label{eq:Ak}
	A_{\LM \ell n} = \sigma_{ \LM \ell \MM n} \, \mathcal{A}_{\ell \MM n} \text{,  and  }\; t = \hatt - \hatt_{\mathrm{ref}} \,,
\end{align}
where in \eqn{eq:Ak}, $\sigma_{ \LM \ell \MM n}$ are the mixing coefficients between spherical and spheroidal harmonics \cite{Berti:2014fga,London:2018nxs}, and $\hatt$ is an observer's time coordinate.
\par Three conventions are used to practically define the ringdown region and assist model construction.
First, we note that different choices of reference time, $\tref$, result in different values of $\Ak$ for different \bbh{} configurations.
That is, when comparing two such conventions, for example, $\tref^{(a)}$, defined at the peak of $|\frac{d}{dt}{h}_{22}|$ (as in Refs.~\cite{Kamaretsos:2012bs,Gossan:2011ha}),
and $\tref^{(b)}$ defined at the peak of $h_{22}$ (as is used here), the resulting values of $\Ak$ differ according to
\begin{align}
	\label{eq:tref}
	\Ak^{(a)}/\Ak^{(b)} \;\; \propto \;\; e^{ \, i \cw_{\lmn} ( \; \tref^{(b)} \, - \,\tref^{(a)} \; ) \, }.
\end{align}
As the \qnm{} frequencies and decay times depend nontrivially on the initial \bh{} masses and spins, \eqn{eq:tref} communicates that different conventions for $\tref$ generally result in different pictures of \qnm{} excitation.
\fig{fig:tdfd} illustrates the effect for the dominant \qnm{} amplitude, $\mathcal{A}_{220}$, on the space of initially nonspinning \bbh{}s with masses $M_1$ and $M_2$, with a symmetric mass ratio, $\eta = M_1 M_2 / (M_1+M_2)^2$.
Note that $\sigma_{22220}\approx 1$ for all remnant \bh{} spins, thus $\mathcal{A}_{220} \approx A_{2220}$ \cite{Berti:2014fga,London:2018nxs}.
Here, not only does the choice of $\tref$ affect $\mathcal{A}_{220}$'s functional form, but $\tref^{(b)}$ (at peak $h_{22}$) is revealed to be a remarkably simple choice, one resulting in $|\mathcal{A}_{220}| \approx 4\eta$.
For this reason, $\tref=\tref^{(b)}$ is used here.
\par Second, we note that throughout the nonspinning \bbh{} parameter space, system mass and angular momentum typically continue to evolve prior to $t \approx 16~M$, and are constant thereafter~\cite{Buonanno:2006ui,Husa:2015iqa,London:2014cma}.
Thus, initial work considered the start of ringdown to effectively being $16~M$ after $\tref^{(b)}$ \cite{Kamaretsos:2012bs,Gossan:2011ha}.
To accommodate progenitor \bh{}s with high spins, this work considers the start of ringdown to be at $T_\mathrm{start}=\T0~M$ to the right of $\tref^{(b)}$. Ringdown is held to end when the simulation is dominated by numerical noise \cite{London:2014cma}.
\par While both $T_\mathrm{start}$ and $\tref$ relate to the start of ringdown, $\tref$ refers to the intrinsic start of the perturbation, while $T_\mathrm{start}$ relates to the extrinsic choice of which segment in time contains \qnm{} \rd{}. 
Both quantities are relevant in that ringdown is considered to start within the data at $t=T_\mathrm{start}$. 
However, \eqns{eq:hlmn}{eq:Ak} communicate that the dependence of each \qnm{} amplitude on initial parameters is only affected by $\tref$.
\par Lastly, we note that the orbital phase between simulations follows no \textit{a priori} convention near \rd{}. This is overcome by rotating the decomposition frame about the $z$ axis such that $A_{2220}$ is real.

\section{Model Construction}
The \rd{} of each $b$th \nr{} simulation corresponds to an initial parameter list ${\mlam_b = \{M_1,M_2,\chi_s,\chi_a\}}$,
where $M_1>M_2$, $M=M_1+M_2$ and
\begin{align}
	&\eta = M_1 M_2 / M^2,\;\; \delta=\sqrt{1-4\eta},  \\ \nonumber
	&\chi_s = (M_1 \chi^{(z)}_1 + M_2 \chi^{(z)}_2)/M, \\ \nonumber
	&\chi_a = (M_1 \chi^{(z)}_1 - M_2 \chi^{(z)}_2)/M \; .
\end{align}
Here, the dimensionless spin, $\chi_j^{(z)}$, is the $j^{\mathrm{th}}$ \bh{} spin's z component divided by $M_j^2$.
As in Ref.~\cite{London:2014cma}, each simulation's $\Ak{}$ is determined numerically using least-squares regression in the frequency domain.
The system's initial parameters are related to its remnant's mass and spin via phenomenological fitting formulas~\cite{Frauendiener:2011zz,Healy:2016lc,Jimenez-Forteza:2016oae}.
Given the resulting \qnm{} content, the fit is reapplied over validating fitting regions with ${T_\mathrm{start}}\rightarrow{T_\mathrm{start}}'$ on $[T_\mathrm{start}, T_\mathrm{start}+10~M]$.
This enables the identification of \textit{incidental} \qnm{s} which do not satisfy time translational symmetry (i.e. $\Ak{}$ varies significantly with ${T_\mathrm{start}}'$, when it should be constant).
While not physical, these incidental \qnm{s} capture information that can be attributed either to the pre-\qnm{} regime, or to time dependent numerical noise not of interest for modeling.
%
\par In particular, it is found that the \qnm{s} with $n>0$, while inconsistent with noise, do \textit{not} display time translational symmetry over the nonprecessing parameter space.
Thus, the median over the validation regions of only the nonincidental $\Ak$ is stored for modeling.
\def\params{\eta,\delta,\chi_s,\chi_a}
%
%
%
\par The desired \qnm{} amplitude model, $\Ak(\mlam)$, interpolates over $\{\mlam{} \rightarrow \Ak{}\}_b$. 
Each $\Ak(\mlam)$ is found to be well represented by a post-Newtonian-like expansion:
$ A_{k}(\mlam) = \eta \sum_u a_{uk} C_u(\mlam) $,
where each $C_u(\mlam)$ represents a unique product of $\mlam$'s elements to some power
(e.g. $C_u \in \{1,\eta,\chi_s,\eta\chi_s,\eta^2, ...\}_u$), and $k$ encodes $(\bar{\ell},\bar{m},\ell,n)$~\cite{Blanchet:2013haa,London:2014cma}.
From this perspective, determining each $A_{k}(\mlam)$ is equivalent to finding each $a_{uk}$.
As this problem is linear in $C_u(\mlam_b)$, $a_{uk}$ are determined using least-squares multinomial regression \cite{London:2018nxs}.
%
\section{Results}
%
Each $\Ak(\params)$ is shown in \eqns{eq:A22_fit_1}{eq:A43_fit_2}.
Residuals for each fit are found to be approximately Gaussian, zero centered, with an average standard deviation of \check{4.66\%} in both real and imaginary parts.
%
%
These results enable the evaluation of \rdnp{} according to
\begin{align}
	\label{eq:ht}
	h(r,t,\iota,\phi) = \frac{GM}{ r c^2} \, \sum_{\LL,|\MM|\le \LL} \sum_{\ell n} \, \Ak \, e^{i\cw_{\ell \MM n}t} \, \Yj{} \; .
\end{align}
\eqn{eq:ht} is limited to the indices present in \eqns{eq:A22_fit_1}{eq:A43_fit_2} with the exception that nonprecessing symmetry yields $\MM<0$ terms from $h_{\LL,-\MM} = (-1)^{\LL} h^{*}_{\LL,\MM}$ \cite{Blanchet:2013haa}.
As in previous studies, additional mode amplitudes, such as those with $\MM=0$, are not modeled as they are known to not significantly contribute to the overall \gw{} emission~\cite{Kamaretsos:2011um, London:2017bcn,London:2014cma}.
%
%
%
\par \fig{fig:results} displays select \qnm{} amplitudes, phases, as well as model validation.
The left panel of Fig.~\ref{fig:results} compares calibration points (colored circles) with the model for $|A_{2220}|$ over the $(\eta,\chi_s)$ parameter space.
Color differences between calibration points and the model's smooth gradient correspond to noise within the calibration set.
%
%
%
%
%
\par As in Ref.~\cite{London:2014cma}, $A_{3230}$ and $A_{4440}$ are found to have nonmonotonic amplitudes which correspond to rapid and localized changes in relative phase, $\arg(\Ak)$. Due to nonprecessing symmetry, these are the strongest subdominant \qnm{}s for equal-mass \bbh{}s~\cite{Blanchet:2013haa, Cotesta:2018fcv}.
%
%
In the central panel of Fig~\ref{fig:results}, we see for the first time that these abrupt transitions in phase are a robust feature of the nonprecessing parameter space.
\section{Model Validation} The right panel of Fig.~\ref{fig:results} shows validation of \rdnp{} against \check{5} select non-calibration \nr{} waveforms from the {BAM} code~\cite{Bruegmann:2006at, Husa:2007hp}.
\begin{figure*}[ht]
	\begin{tabular}{cc}

		\includegraphics[width=0.44\textwidth]{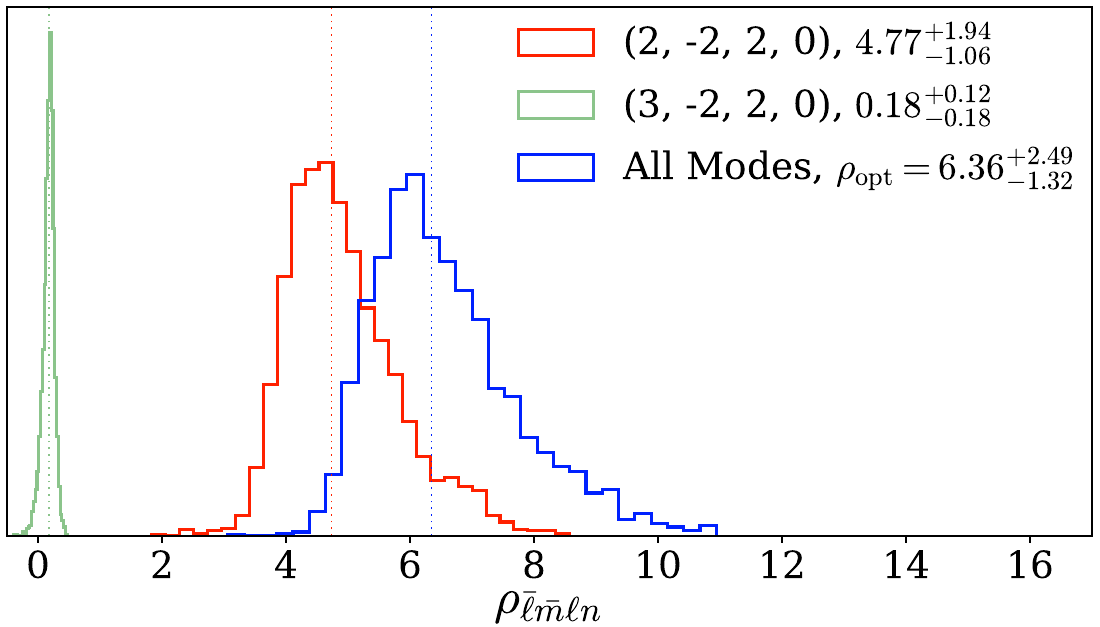} &
		\includegraphics[width=0.455\textwidth]{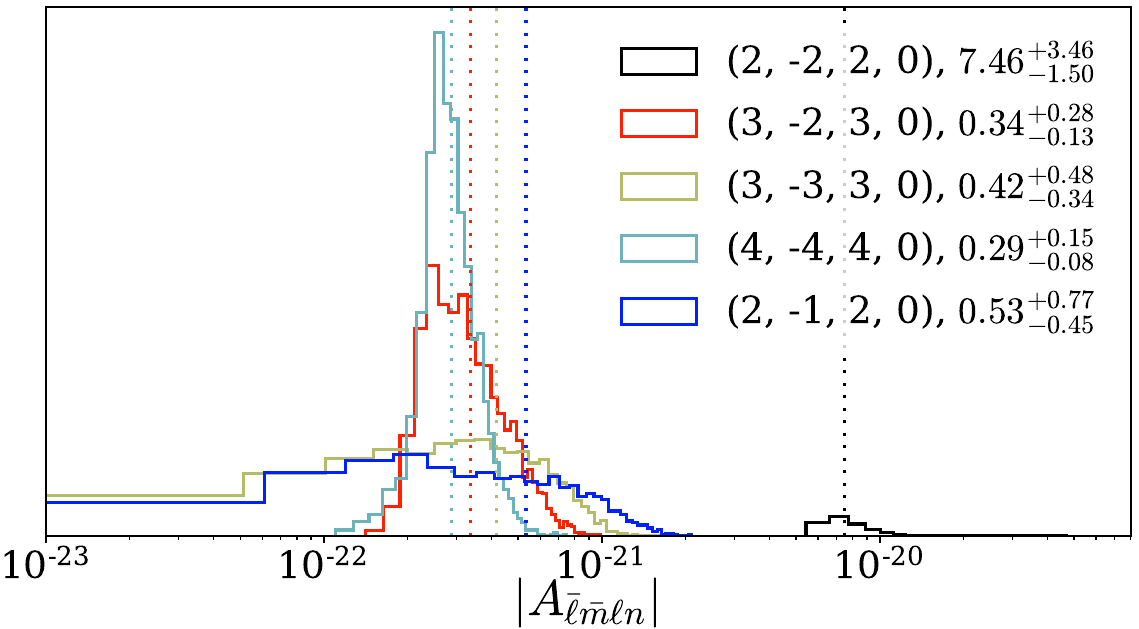}

	\end{tabular}
	\caption{Posterior sample postprocessing for \check{GW150914}: All panels show normalized posterior distributions, with related medians and $90\%$ credible intervals in legends.
	Median values are shown with dotted lines.
	{Left}: Signal to noise ratio (SNR) attributed to each \qnm{} with indices $(\ell,\MM,n)$ within a spherical multipole with indices $(\LL,\MM)$ formatted as $(\LL,\MM,\ell,n)$.
  The attributed SNR for the two most significant \qnm{s} are shown.
	The optimal SNR, which is the \gr{} prediction for the total \rd{} SNR, is shown for reference.
	%
	%
	Right: \gr{} predictions for absolute \qnm{} amplitudes via \eqns{eq:A22_fit_1}{eq:A43_fit_2}. Here medians and credible intervals are scaled by $10^{21}$.
	%
	%
	}
	\label{fig:discuss}
\end{figure*}
Here, \nr{} \rd{} plays the role of a hypothetical signal at inclination $\iota$, and \rdnp{} plays the role of a template at the same inclination, with independent polarization and orbital phase.
The normalized inner product, or match, $( h_{\mathrm{RDNP}} | h_{\mathrm{NR}} )$, is weighted by the anticipated \aligo{} zero-detuned noise power spectrum at design sensitivity ~\cite{advLIGOcurves} and calculated following Eq.~(46) of Ref.~\cite{Harry:2016ijz}, with a starting frequency $f_{\mathrm{min}} = 30$~Hz for the integral.
\rdnp{} is evaluated at the same intrinsic parameters as the \nr{} waveform such that there are no spin components within the orbital plane.
\rdnp{} matches extremely well with \nr{} cases in and out of the calibration region, often having matches above 0.998.
This is the case even for a precessing waveform, ``q1.2\_p'', similar to \check{GW150914}~\cite{TheLIGOScientific:2016wfe}.
Not shown is the high spin aligned validation case $(M_1/M_2=8,\chi_s=0.85)$.
The nonlinear regime for this system extends to approximately $40~M$. When taking this into account, \rdnp{} matches as low as \check{0.97} for $\iota \approx \pi/2$, but \check{0.99} and well above for \check{$|\iota-\pi/2| > \pi/6$}.
%
%
%
%
%
\section{Discussion}
%
\par \rdnp{} has been presented to model the \rd{} of nonprecessing \bbh{} systems.
While \rdnp{} matches well with \nr{} simulations, there are multiple avenues for improvement.
\rdnp{} does not model precession.
\rdnp{} also does not model the apparent nonlinear \qnm{s} reported in Ref.~\cite{London:2014cma}. This may be most important for systems with high aligned spins, where the nonlinear regime is extended.
\par \rdnp{} provides redundant mode information. For example, via \eqn{eq:Ak}, $A_{32320}$ and $A_{32220}$ differ only by factors of $\sigma_{\LM\lmn}$. This allows \rdnp{}'s consistency with perturbation theory to be quantified by comparing modeled ratios of $\sigma_{\LM\lmn}$ to perturbation theory predictions~\cite{Kelly:2012nd,London:2014cma}.
For \rdnp{}, ratios of $\sigma_{\LM\lmn}$ agree with perturbation theory within $5\%$ in amplitude and $15\%$ in phase. This agreement could be improved in future treatments.
\par Future \rd{} models should be calibrated to a larger set of more accurate \nr{} simulations. Like Ref.~\cite{Baibhav:2017jhs}, the current work is limited by quality concerns between simulations of different numerical codes.
\rdnp{} and related techniques may of be of use in constructing \nr{}-tuned full signal models with accurate mergers.
In particular, \rdnp{} may be interfaced directly with the analytic merger-ringdown ansatz proposed in Ref.~\cite{McWilliams:2018ztb}.
Primarily, it is expected that \rdnp{} may be of use aiding tests of \gr{} during LIGO's third observing run. In that setting, many practical questions regarding \rd{} are pertinent.
%
\subsection{Data analysis example }

%
\par The following questions are briefly considered: How much SNR is in ringdown? How much SNR can be attributed to subdominant \qnm{s}? Can the \qnm{} amplitudes be constrained, and can their relative phases?
To proceed, \rdnp{} is applied to inferred posteriors of \check{GW150914}'s parameters via a higher-multipole \imr{} model, {\sc{PhenomHM}} \cite{London:2017bcn}.
%
%
Here, {\sc{PhenomHM}} is applied to the Bayesian inference of GW150914~\cite{Vallisneri:2014vxa}, according to Ref.~\cite{TheLIGOScientific:2016wfe}, and then posterior samples are input to \rdnp{} to yield GR predictions for quantities reported in \fig{fig:discuss}.
This approach yields GR predictions independently of individual mode \snr{}~\cite{Berti:2005ys}.
\par The top panel of Fig.~\ref{fig:discuss} shows the posterior distribution for $\rho_{\mathrm{opt}}$, the estimated total \rd{} SNR (red) for \check{GW150914} \cite{Usman:2015kfa,Finn:2000hj}.
Note that the approximately face-off nature of \check{GW150914} means that $m<0$ \qnm{}s are prevalent~\cite{TheLIGOScientific:2016wfe}.
Additional posteriors are shown for the \snr{} contributed by a single \qnm{}
$$
	\rho_{\LL \MM \ell n}=\rho_{\mathrm{opt}} - \rho_{\mathrm{opt|\LL \MM \ell n}}\;,
$$
where the single interferometer $\rho_{\mathrm{opt| \LL \MM \ell n}}^2$ is the inner product between a \rdnp{} evaluation with all \qnm{s}, and without the $(\LL,\MM,\ell,n)$ mode.
Not surprisingly, the majority of the ringdown SNR can be attributed to the dominant quadrupole, $A_{2-220}$.
\def\DSNRHM{1.63^{+0.86}_{-0.72}}
\par Intriguingly, \fig{fig:discuss}'s suggests that the total amount of SNR attributed to subdominant modes is of order 1. It is found that the distribution of SNR attributed to the subdominant modes indeed yields \check{$$\rho_\mathrm{opt}-\rho_{2-220}=\DSNRHM\;.$$}
This \check{order 1 contributed SNR} is explained by $\rho_{\mathrm{opt}}^2$ having cross terms that are proportional to $\rho_{2-220}\,\rho_{\LL \MM \ell n}$, meaning that the larger $\rho_{2-220}$, the larger the effect of $\rho_{\LL \MM \ell n}$ on $\rho_{\mathrm{opt}}$.
%
%
\par The right panel of Fig.~\ref{fig:discuss} shows GR predictions for \qnm{} amplitudes.
For the first time it can be seen that \qnm{}s with odd $m$ have amplitudes which are difficult to constrain.
Equations~(\ref{eq:A22_fit_1})-(\ref{eq:A43_fit_2}), along with well known difficulty measuring component spins (e.g. \cite{Purrer:2015nkh}), yield a straightforward explanation: uncertainty in $A_{\LL \MM \ell n}$ with odd $m$ is dominated by uncertainty in the component spins.
%
%
Detector networks with greater sensitivity and more interferometers may overcome this limitation~\cite{Purrer:2015nkh}.
%
%
%
%
\par The amount of SNR attributed to higher \qnm{s} and the possibility of using the \qnm{} amplitudes and the relative phase of \eqns{eq:A22_fit_1}{eq:A43_fit_2} to test \gr{} illuminates a need to further development of analysis pipelines.
Much work has been done in this regard (e.g.~\cite{Gossan:2011ha,Bhagwat:2016ntk}), and the interface of \rdnp{} with existing pipelines is ongoing ~\cite{LALSuite}.
%
\subsection{Informing analytic ringdown }
%
While this work develops a numerical representation for \rd{}, an analytic theory linking \rd{} excitations to the initial binary is at present nonexistent.
\par It may be postulated that such a theory requires physical choices about an observer's time coordinate relative to features in the full \gw{} signal.
The work presented here strongly suggests that the natural reference time for \rd{} is near the peak strain.
This result is observationally convenient, given the amount of signal power in that regime.
However, it is also counterintuitive, as the peak strain resides in a nonperturbative regime, where the remnant \bh{} is not Kerr~\cite{Bhagwat:2016ntk}.
%
%
\par It is also apparent in this work that any potential analytic theory of \rd{} excitation must reproduce the abrupt transitions in phase seen in \fig{fig:results}.
These transitions have no counterpart in the treatment of inspiral, despite the fact that, the lowest order scaling of \pn{} and \qnm{} amplitudes appears to be identical~\cite{London:2014cma}.
%
\par In total, the results presented here may not only aid tests of \gr{}, but they may also help motivate and constrain an analytic (\pn{}-like) theory of \rd{}, one in which \rd{} amplitudes are linked to the nonperturbative regime, yet manifestly consistent with linear \pt{}.
\vspace{-1cm}
\begin{widetext}
\begin{align}
  \label{eq:A22_fit_1}
  A_{2220}  \; &= \; \eta \, ( \, -0.6537\,\chi_s \; + \; -4.0071 \; ) \\
	\label{eq:A21_fit_1}
	A_{2120}  \; &= \; \eta \, ( \, 2.3488\,e^{2.6631i}\,\delta \; + \; (0.8011\,e^{5.7070i})\,\chi_a
	+ \; (3.5828\,e^{5.5223i})\,\eta\,\delta \; + \; (1.1774\,e^{0.4254i})\,\chi_s\,\delta
	\\ \nonumber & \quad  \;
	 + \; (0.6260\,e^{5.3457i})\,\chi_s\,\chi_a \; ) \\
	\label{eq:A33_fit_1}
	A_{3330}  \; &= \; \eta \, ( \, 2.6412\,e^{2.9880i}\,\delta \; + \; (1.6030\,e^{0.6655i})\,{\delta}^{2}
	 + \; (1.0354\,e^{3.6096i})\,\chi_s\,\delta \; + \; (0.4911\,e^{4.7347i})\,{\chi_a}^{2} \; )
	\\
	%
	\label{eq:A32_fit_1}
	A_{3230}  \; &= \; \eta \, ( \, 2.5707\,e^{4.1427i}\,\eta \; + \; (9.4216\,e^{0.8076i})\,{\eta}^{2}
	 + \; (0.5973\,e^{2.1816i})\,\eta\,\chi_s \; + \; (0.2104\,e^{4.9043i})\,{\chi_a}^{2}
	\\ \nonumber & \quad  \;
	+ \; (0.4417\,e^{5.4544i})\,\chi_a\,\delta \; + \; (0.9439\,e^{1.7614i})\,{\delta}^{2} \; )
	\\
	\label{eq:A32_fit_2}
	A_{3220}  \; &= \; \eta \, ( \, 1.3407\,e^{2.9466i}\,\eta \; + \; (0.0717\,e^{5.5304i})
	 + \; (0.1061\,e^{2.6432i})\,{\chi_s}^{2} \; + \; (0.9894\,e^{2.9294i})\,\eta\,\chi_s
	 + \; (0.3735\,e^{3.3290i})\,\chi_a\,\delta \; )
	  \\
		%
	\label{eq:A44_fit_1}
	A_{4440}  \; &= \; \eta \, ( \, 1.3284\,e^{2.6831i}\,{\delta}^{2} \; + \; (1.1619\,e^{0.4142i})\,{\delta}^{3}
	 + \; (1.2790\,e^{4.7226i})\,\chi_s\,{\chi_a}^{2}\,\delta \; + \; (1.2387\,e^{4.5616i})\,\chi_s\,{\chi_a}^{3}
	\\ \nonumber & \quad  \;
	 + \; (1.2909\,e^{2.8120i})\,\chi_s\,{\delta}^{3} \; + \; (42.3575\,e^{6.1418i})\,{\eta}^{4} \; ) \\
	\label{eq:A43_fit_1}
	A_{4330}  \; &= \; \eta \, ( \, 0.0411\,e^{2.6441i}\,\chi_a \; + \; (0.0486\,e^{3.2085i})\,{\chi_s}^{2}
	 + \; (0.8078\,e^{2.7461i})\,\eta\,\delta \; + \; (0.1940\,e^{3.0292i})\,\chi_s\,\delta
	\\ \nonumber & \quad  \;
	 + \; (0.0529\,e^{3.5830i})\,{\chi_a}^{2} \; + \; (0.0358\,e^{0.1731i})\,{\delta}^{2} \; )  \\
	\label{eq:A43_fit_2}
	A_{4340}  \; &= \; \eta \, ( \, 0.5665\,e^{3.3992i}\,\delta \; + \; (0.1457\,e^{4.7476i})\,\chi_a
	 + \; (0.8239\,e^{1.8174i})\,\eta\,\chi_a \; + \; (0.0507\,e^{4.7495i})\,\chi_s\,\chi_a
	\\ \nonumber & \quad  \;
	 + \; (0.9806\,e^{0.6029i})\,{\delta}^{3} \; + \; (10.1678\,e^{6.2185i})\,{\eta}^{2}\,\delta \; )
\end{align}
\end{widetext}
%
\section{Acknowledgements}

The author thanks Mark Hannam, Sascha Husa and Scott Hughes for useful discussions and access to their numerical waveforms.
Additional thanks are extended to the LIGO-Virgo collaboration for their comments and review.
This research has made use of data, software and/or web tools obtained from the LIGO Open Science Center, a service of LIGO Laboratory, the LIGO Scientific Collaboration and the Virgo Collaboration. LIGO is funded by the U.S. National Science Foundation. Virgo is funded by the French Centre National de Recherche Scientifique (CNRS), the Italian Istituto Nazionale della Fisica Nucleare (INFN) and the Dutch Nikhef, with contributions by Polish and Hungarian institutes.
The work presented in this paper was supported by Science and Technology Facilities Council (STFC)
Grant No. ST/L000962/1, European Research Council Consolidator Grant No. 647839, Spanish Ministry of Economy and Competitiveness Grants No. CSD2009-00064, No. FPA2013-41042-P and
No. FPA2016-76821-P, the Spanish  Agencia Estatal de Investigaci\'{o}n, European Union FEDER funds, Vicepresid\`{e}ncia i Conselleria d'Innovaci\'{o}, Recerca i Turisme, Conselleria d'Educaci\'{o},  i Universitats del Govern de les Illes Balears, and the Fons Social Europeu.
{\sc BAM} simulations were carried out at Advanced Research Computing (ARCCA) at Cardiff, as part of the European PRACE petascale computing initiative on the clusters Hermit, Curie and SuperMUC,
on the U.K. DiRAC Datacentric cluster and on the BSC MareNostrum computer under PRACE and RES (Red Espa\~{n}ola de Supercomputaci\'{o}n) allocations.
\definecolor{UrlColor}{rgb}{0.75, 0, 0}
\bibliographystyle{apsrev4-1}
\bibliography{np.bib}
\end{document}